\documentclass[final]{pasj00}
\Received{2013 September 24}
\Accepted{2014 March 3}
\Published{$\langle$---$\rangle$}
\SetRunningHead{Akahori, Kumazaki, Takahashi, and Ryu}{Exploring the IGMF by Faraday Tomography}
\usepackage{times}

\begin{document}

\title{Exploring the Intergalactic Magnetic Field by Means of Faraday Tomography}
\author{
Takuya \textsc{Akahori}$^{1}$, Kohei \textsc{Kumazaki}$^2$, Keitaro \textsc{Takahashi}$^3$, and Dongsu \textsc{Ryu}$^4$
}
\affil{
$^1$Sydney Institute for Astronomy, School of Physics, The University of Sydney, NSW 2006, Australia, akahori@physics.usyd.edu.au\\
$^2$University of Nagoya, 2-39-1, Kurokami, Kumamoto 860-8555, kumazaki@a.phys.nagoya-u.ac.jp \\
$^3$University of Kumamoto, 2-39-1, Kurokami, Kumamoto 860-8555, keitaro@sci.kumamoto-u.ac.jp\\
$^4$Department of Astronomy and Space Science, Chungnam National
University, Daejeon, Republic of Korea, ryu@canopus.cnu.ac.kr
}

\maketitle

\begin{abstract}
Unveiling the intergalactic magnetic field (IGMF) in filaments of galaxies is a very important and challenging subject in modern astronomy. In order to probe the IGMF from rotation measures (RMs) of extragalactic radio sources, we need to separate RMs due to other origins such as the source, intervening galaxies, and our Galaxy. In this paper, we discuss observational strategies for the separation by means of Faraday tomography (Faraday RM Synthesis). We consider an observation of a single radio source such as a radio galaxy or a quasar viewed through the Galaxy and the cosmic web. We then compare the observation with another observation of a neighbor source with a small angular separation. Our simulations with simple models of the sources suggest that it would be not easy to detect the RM due to the IGMF of order $\sim 1~{\rm rad~m^{-2}}$, an expected value for the IGMF through a single filament. Contrary to it, we find that the RM of at least $\sim 10~{\rm rad~m^{-2}}$ could be detected with the SKA or its pathfinders/precursors, if we achieve selections of ideal sources. These results would be improved if we incorporate decomposition techniques such as RMCLEAN and QU-fitting. We discuss feasibility of the strategies for cases with complex Galactic emissions as well as with effects of observational noise and radio frequency interferences.
\end{abstract}

\section{Introduction}
\label{section1}

The intergalactic medium (IGM) in the cosmic web of filaments and clusters of galaxies is expected to be permeated with the intergalactic magnetic field (IGMF) (\cite{wrsst11,rsttw11}). The IGMF plays crucial roles in various subjects of astrophysics; propagations of ultra-high energy cosmic-rays and $\gamma$-rays (\cite{mur08,das08,tnys09,rdk10,tak08,tak11,tak12}), radio emissions in galaxy clusters (\cite{fer12,fo13}), substructures during cluster mergers (\cite{asa05,taki08}), and configurations of magnetic fields in spiral galaxies (\cite{smk10}). Seed IGMFs could be generated in inflation, phase transition, and recombination eras (\cite{gr01,tak05,ich06}), during cosmic reionization (\cite{gfz00,lap05,xu08,ads10}), and through cosmological shock waves (\cite{kcor97,rkb98}). The seed fields of any origins could be further amplified through compression and turbulence dynamo in the structure formation (\cite{rkcd08,dbd08,cr09,sbsa10}). Also, leakages of magnetic fields and cosmic rays from galaxies should be taken into consideration (\cite{ddlm09,mb11}).

The above diverse processes underline the importance of observational tests for the IGMF. One of a few possible methods to probe cosmic magnetic fields is to utilize Faraday rotation in radio polarimetry (\cite{ct02,gbf04,beck09,gov10}). A rotation of the polarization angle is proportional to the square of the wavelength, and the proportionality constant, rotation measure (RM), provides an integration of magnetic fields with weights of electron densities. This conventional method, however, could work only in the case of observing a single polarized radio source. Otherwise, in cases of multiple emitters along the line-of-sight (LOS), a rotation of the polarization angle draws a complex curve (\cite{bd05}, hereafter BD05), and RM cannot be easily estimated. Moreover, RMs of a few to several tens ${\rm rad~m^{-2}}$ are usually associated with radio sources (\cite{sc86,osu12}) and the Galaxy (\cite{mao10,opp12}). These RMs are larger than expected RMs through filaments, $\sim 1-10~{\rm rad~m^{-2}}$ (\cite{ar10,ar11,agr13}), and cannot be easily separated from an observed RM by the conventional method. Therefore, we need to establish alternative methods which allow us to estimate hidden RM components along the LOS.

As a method to separate multiple sources and RMs along the LOS, a revolutionary technique, called Faraday RM synthesis or Faraday tomography\footnote{We consider one-dimensional reconstruction in this paper. Although the phrase ``tomography" is generally used as an attempt to reconstruct the actual 3D distribution from observed integrals through the volume, we call this technique as Faraday tomography throughout this paper, foreseeing future 3D imaging of the cosmic web.}, was first proposed by \citet{burn66} and extended by BD05. Previous works for the interstellar medium (\cite{skb07,skb09}), the Galaxy (\cite{mao10}), external galaxies (\cite{hbe09}), and active galactic nuclei (\cite{osu12}) have demonstrated that the technique is powerful to resolve RM structures along the LOS. It would be thus promising to study the IGMF in eras of wide-band radio polarimetry including Square Kilometer Array (SKA) and its pathfinders/precursors such as Low Frequency Array (LOFAR), Giant Meterwave Radio Telescope (GMRT), and Australia SKA Pathfinder (ASKAP) (see \cite{bfss12}, a summary of telescopes therein). 

In this paper, we discuss observational strategies to probe the IGMF by means of Faraday tomography. We consider frequency coverages and numbers of channels of future observations. Faraday tomography is in general improved by incorporating decomposition techniques such as RMCLEAN (\cite{hea09}) and QU-fitting (\cite{osu12}). However, decomposition techniques have their own uncertainties (\cite{far11,kum13}). Therefore, we concentrate on a standard method of Faraday tomography without any corrections to see its original potential. In fact, the decomposition is powerful for our study, which was addressed in a separate paper (\cite{ide14a}). The rest of this paper is organized as follows. In section 2, we introduce the method of Faraday tomography and describe our model. The results are shown in section 3, and the discussion and summary follow in section 4 and 5, respectively.

\section{Model and Calculation}
\label{section2}

\subsection{Faraday Tomography}
\label{section2.1}

We first summarize a basic concept of Faraday tomography following a manner described by BD05. The readers who have interests in detailed derivations and improvements of algorithm should refer to recent works (\cite{hea09,fssb11,lbcd11,ast11}).

A fundamental observable quantity is the complex polarized intensity, $P(\lambda^2) = Q(\lambda^2)+iU(\lambda^2)$, at a given wavelength, $\lambda$, where $Q$ and $U$ are the Stokes parameters. $P(\lambda^2)$ is given by an integration of intensities along the LOS as (\cite{burn66}),
\begin{equation}\label{eq1}
P(\lambda^2) 
=\int_{-\infty}^{\infty} F(\phi,\lambda^2) e^{2i\phi \lambda^2} d\phi,
\end{equation}
where $F$ is the Faraday dispersion function (FDF) which is a complex polarized intensity at a given Faraday depth, $\phi$. Faraday depth is defined as
\begin{equation}\label{eq2}
\phi(x)=0.81 \int_{x}^{0}n_{\rm e}(x')B_{||}(x')dx'
\end{equation}
in units of ${\rm rad~m^{-2}}$, where $n_{\rm e}$ is the thermal electron density in ${\rm cm^{-3}}$, $B_{||}$ is the LOS magnetic field strength in ${\rm \mu G}$, 
and $x'$ is the LOS physical distance in pc.

The FDF should be a function of $\lambda$ in general (BD05), but, following \citet{burn66}, we assume that all considerable radio sources have similar spectra and neglect wavelength dependence of the FDF. Hence equation (\ref{eq1}) has the same form as the Fourier transform and the FDF can be formally obtained by the inverse Fourier transform:
\begin{equation}\label{eq:F}
F(\phi)=\int_{-\infty}^{\infty} P(\lambda^2)e^{-2i\phi \lambda^2} d\lambda^2.
\end{equation}
However, this inversion is not practically perfect, because $P(\lambda^2)$ is not defined for negative $\lambda^2$ and the coverage for positive $\lambda^2$ is limited in real observations.

BD05 generalized equation (\ref{eq:F}) by introducing a window function, $W(\lambda^2)$, where $W(\lambda^2)=1$ if $\lambda^2$ is in observable bands, otherwise $W(\lambda^2)=0$. Observed complex polarized intensity can then be written as 
\begin{equation}\label{eq:tildeP}
\tilde{P}(\lambda^2)
= W(\lambda^2)P(\lambda^2)
= W(\lambda^2)\int_{-\infty}^{\infty} F(\phi) e^{2i\phi \lambda^2} d\phi,
\end{equation}
where the tilde indicates observed or reconstructed quantities. From $\tilde{P}(\lambda^2)$, we can obtain an approximate reconstruction of the FDF as,
\begin{equation}\label{eq:tildeF}
\tilde{F}(\phi)
= \int_{-\infty}^{\infty} \tilde{P}(\lambda^2) e^{-2i\phi \lambda^2}
  d\lambda^2.
\end{equation}

We define the rotation measure spread function (RMSF):
\begin{equation}\label{eq:RMSF}
R(\phi)=K\int_{-\infty}^{\infty} W(\lambda^2) e^{-2i\phi \lambda^2} d\lambda^2,
\end{equation}
where
\begin{equation}\label{eq:K}
K=\left(\int_{-\infty}^{\infty} W(\lambda^2) d\lambda^2 \right)^{-1},
\end{equation}
is the normalization. Applying the convolution theorem, the approximate FDF, $\tilde{F}(\phi)$, can be written as,
\begin{equation}\label{eq:convolution}
\tilde{F}(\phi)
= K^{-1} F(\phi)\ast R(\phi)
= K^{-1} \int_{-\infty}^{\infty} F(\phi-\phi') R(\phi') d\phi'.
\end{equation}
Therefore, the reconstruction (\ref{eq:convolution}) is perfect only if the RMSF reduces to the delta function, $R(\phi)/K \rightarrow \delta(\phi)$, for a (unphysical) complete observation, i.e. $W(\lambda^2) = 1$ for all $\lambda^2$ (Eq. \ref{eq:RMSF}). In fact, the RMSF has a finite width (see Figure \ref{f3}), so that the data sampling is incomplete and the reconstruction is not perfect. Equations (\ref{eq:RMSF})-(\ref{eq:convolution}) indicate that the quality of the reconstruction primarily depends on the window function. That is, a wider coverage in $\lambda^2$ space improves the reconstruction, as expected from the analogy with Fourier transform. 

In this paper, we follow the above window-function approach considering the observable bands of SKA and its pathfinders. For other approaches, where the values of non-observed $P(\lambda^2)$ are estimated by some assumptions on the properties of the sources (see \cite{burn66,gw66}; BD05).

\subsection{Promising Target}
\label{section2.2}

Since the sign of $B_{||}$ in equation (\ref{eq2}) can be changed, $\phi(x)$ is not a monotonic function of $x$ in general. Therefore, the FDF, i.e., the distribution of radio sources in $\phi$ space, does not simply indicate the distribution of radio sources in $x$ space. Nevertheless, the FDF is enough to probe the IGMF, because the IGMF can be identified in $\phi$ space as demonstrated in this paper.

We first introduce general behavior of the FDF. If thermal electrons inducing RMs co-exist with cosmic-ray electrons emitting synchrotron polarizations, the FDF has finite thickness in $\phi$ space. Also, an accumulation of emissions and RMs within a source results in a thickness, if the sign of $B_{||}$ changes many times and the accumulation behaves like a random walk process. Such a thickness would be natural for the Galaxy and extragalactic radio sources. In addition, if there are magneto-ionic media in front of radio-emitting region, e.g., associated media such as clouds, H$\alpha$ filaments, and swept IGM by jets, the FDF shifts in $\phi$ space. The shift is also caused by RMs of discrete intervening galaxies (\cite{wpk84,ber12,ham12}) and foreground IGMFs in clusters/filaments of galaxies.

We probe the IGMF in filaments of galaxies from the shift of the FDF in $\phi$ space. Thus, the shifts caused by other origins are contaminations. LOSs containing significant RMs of other origins could be avoided as follows. For removing RMs of galaxy clusters, we could exclude sources behind galaxy clusters (e.g., Coma cluster, \cite{mao10}). The detection limit of current X-ray facilities is enough to substantially exclude RMs of galaxy clusters (\cite{ar11}). RMs of associated media around sources and RMs of intervening galaxies could have a tight correlation with optical absorber systems and/or could show small fractional polarization due to depolarization (\cite{ber12, ham12}). Thus, we could discard sources with such contaminations. Note that the RM of the IGMF would not affect depolarization, since the IGMF is expected to be smooth enough within the beam size of $\sim$arcsecond (\cite{ar11}). Furthermore, RMs for associated media of distant sources could be small due to a $(1+z)^{-2}$ dilution factor. \citet{ham12} estimated the dilution and claimed that sources at $z=1$ should only contribute a standard deviation of RMs $\sim 1.5-3.75~{\rm rad~m^{-2}}$. Therefore, we assume that we can select sources toward which the shifts of the FDF caused by RMs of galaxy clusters, the associated media, and intervening galaxies are negligible.

\subsection{Strategy A: Compact Source behind Diffuse Source}
\label{section2.3}

\begin{figure}[tp]
\begin{center}
\FigureFile(75mm,40mm){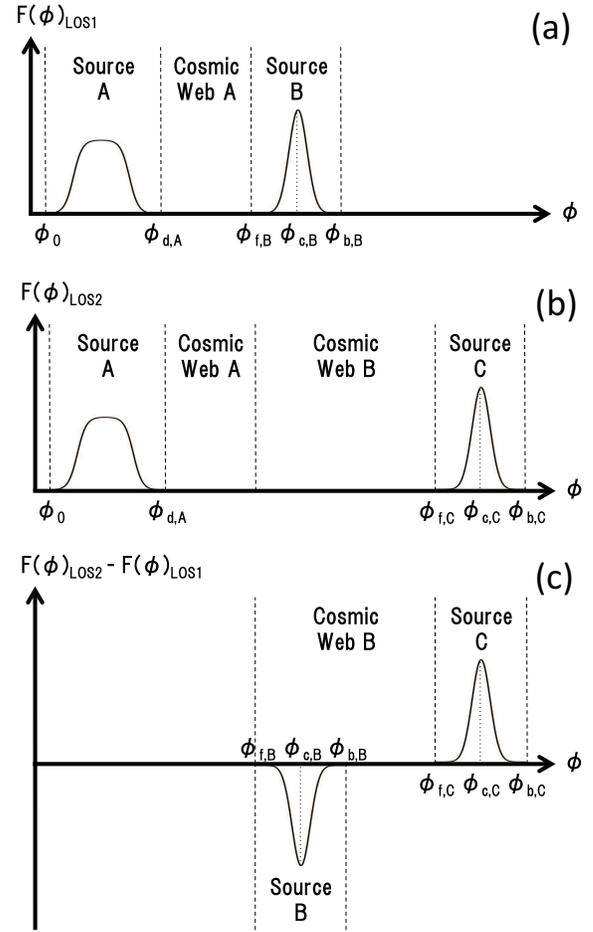}
\end{center}
\caption{(a) Schematic picture of Faraday dispersion function (FDF), $F(\phi)$, in Faraday depth, $\phi$, for ``Strategy A": an observation of a compact source B behind a diffuse source A and the cosmic web A. $\phi_0$ and $\phi_d$ are the positions of the front side and back side edges of the FDF for the diffuse source, respectively. $\phi_f$, $\phi_c$ $\phi_b$ are the positions of the front side, center, and back side edges of the FDF for the compact source, respectively. (b) Same as (a) but for another observation for a compact source C close to the source B on the sky and behind the diffuse source A and cosmic webs A and B. (c) ``Strategy B": difference between two FDFs for two LOSs, ${\rm FDF_{LOS2}}-{\rm FDF_{LOS1}}$.}
\label{f1}
\end{figure}

Hereafter, to make our arguments simple, we mainly describe cases for a pure real $F(\phi)$ obtained if the intrinsic polarization angle is independent of $\phi$. The result for a complex $F(\phi)$ is discussed in Section \ref{section4}.

Using selected sources, we consider possible and suitable FDFs to probe the RM due to the IGMF. Figure \ref{f1}(a) shows an example of the FDF. We consider an observation of a background compact source B (emissions from a radio galaxy or a quasar) viewed through the cosmic web A and a foreground diffuse source A (Galactic emissions). The FDF of the cosmic web A has tiny amplitude, since radio emissions from the IGM are generally much weaker than the others. Therefore, the RM of the cosmic web A can be probed as the ``gap" between FDFs of the two sources. We name this observational strategy as ``Strategy A".

The situation arises if all signs of cumulative RMs of the two sources and the cosmic web coincide. Otherwise, some of them would overlap each other in $\phi$ space. The probability that all signs coincide is $25\%$, which is reasonably high to choose such LOSs from multiple observations. Even in the overlapped case, there would be still the gap if the RM of the cosmic web is much larger than those of sources. Therefore, the probability that we find the gap is expected to be larger than 25~\%.

In this strategy, foreground emissions are necessary to be detected. The intensity of Galactic diffuse emission toward high Galactic latitudes can be scaled as
\begin{equation}\label{eq8}
I \sim 0.95 \left(\frac{f}{1~{\rm GHz}}\right)^{-1.5}
\left(\frac{\Omega}{1~{\rm arcmin^2}}\right)~{\rm mJy}
\end{equation}
($f$ is the frequency and $\Omega$ is the beam size; \cite{gold11}), and is larger toward lower latitudes. The diffuse emission is thus significant, unless we observe very bright compact sources. In section 3.1, we will demonstrate that the diffuse emission is significant even if the background emission is 100-1000 times brighter.

If the source B is a distant source, the FDF of the source B would be Faraday thin (the thickness is small enough, i.e. a delta function). But we keep considering small thickness for compact sources since the thickness as well as the RMSF are notable ambiguities to probe the IGMF, particularly for observations with limited bandwidths. If the source B is Faraday thin, the gap is sharpened and the estimation of the RM due to the IGMF is rather improved.

\subsection{Strategy B: Pair Compact Sources}
\label{section2.4}

Along with Strategy A, we can consider an extended strategy to probe the IGMF. That is, we observe two compact sources and obtain two FDFs (figures \ref{f1}a and \ref{f1}b). We then subtract one FDF from the other to obtain the difference (figure \ref{f1}c). If the directions of two LOSs are close enough, FDFs of the source A and the cosmic web A are cancelled out each other in the difference, then the difference reveals the gap between two FDFs in $\phi$ space. And if the sources B and C are both Faraday thin, the gap precisely indicates the RM of the cosmic web B. We name this observational strategy as ``Strategy B".

\begin{figure}[tp]
\begin{center}
\FigureFile(80mm,40mm){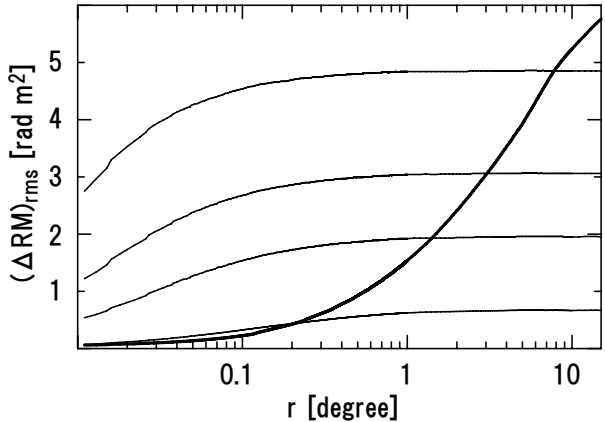}
\end{center}
\caption{
Root-mean-square value of the RM difference, $(\Delta {\rm RM}) _{\rm rms} =\langle \sqrt{{|{\rm RM}(\vec{x}+r)-{\rm RM}(\vec{x})|}^2}\rangle_{\vec{x}}$ averaged over the position $\vec{x}$ of the pixels in the RM map, where $r$ is the separation angle between two pixels. Thick line is for the RM map of the GMF toward high Galactic latitudes (ADPS30, \cite{arkg13}). Thin lines from the bottom to top are for the RM maps of the IGMF (TS0, \cite{ar11}) integrated up to the redshifts of 0.1, 0.3, 0.5, and 1.0, respectively. \label{f2}
}
\end{figure}

Figure \ref{f2} shows how much RM could be different between two LOSs, based on the simulations \citep{ar11,arkg13}. For example, let us suppose that the RM of the cosmic web B is 7.5 ${\rm rad~m^{-2}}$. If the angular separation between the nearby source B and the distant source C is less than $\sim 1^\circ$, the RM difference of the foreground source A could be less than 20\% (1.5 ${\rm rad~m^{-2}}$) of the RM of the cosmic web B. The same level of the RM difference is available for the cosmic web A, if the nearby source B is locate at $z=0.3$ and the angular separation is less than $\sim 0.1^\circ$.

A RM difference of $\sim 1.5$ ${\rm rad~m^{-2}}$ for an angular separation of $\sim 1^\circ$ for the GMF is consistent with the gradient of Galactic RM toward high Galactic latitudes, a few ${\rm rad~m^{-2}}$ per degree (\cite{mao10}). Note that much smaller-scale structures have been observed toward the Galactic plane (\cite{hav06,hav08}). 

\subsection{Prospects}
\label{section2.5}

Let us consider practical cases for strategies A and B. In Strategy A, if the source B is located at $z=1$, RMs of associated media would be $\sim$ a few ${\rm rad~m^{-2}}$. Therefore, if the RM of the cosmic web A exceeds $\sim$ several ${\rm rad~m^{-2}}$, it could be predominant on the shift of the FDF of the source. In Strategy B, if the source B is located at $z=0.1$, the RM difference of the cosmic web A is $\sim 0.3$ ${\rm rad~m^{-2}}$ for the angular separation of $\sim 0.1^\circ$. The source B could contain RMs of $\sim 10$ ${\rm rad~m^{-2}}$ for associated media. Thus, if the RM of the cosmic web B exceeds $\sim 15-20$ ${\rm rad~m^{-2}}$, it could be predominant on the shift of the FDF of the source C. Such a value is likely if the LOS goes through more filaments than the average between the observer and the sources. Moreover, if we can obtain information of optical absorptions and/or depolarization and can select the sources toward which RMs of associated media and intervening clouds are insignificant, we could identify a much smaller RM of the cosmic web B.

We expect that there will be a number of ideal sources for Strategies A and B in millions of candidates to be found in future observations. For example, no absorber systems were detected toward about a half of 84,534 quasars so far \citep{zhu13} and a number of radio sources show large fractional polarization \citep{ham12}. Pair sources are still rare in the largest catalog with the average separation $\sim 1^\circ$ (\cite{tss09}). But the separation will decrease by $\sim 0.1^\circ$ with SKA precursors and $\sim 0.01^\circ$ with the SKA, so that the number of the pairs would soon increase dramatically. 

Supposing that optical observations confirm two independent sources B and C, Strategy B could be available even if the sources are in a single beam and are not resolved in space. But if the Faraday depth between two sources is less than the full width at half maximum (FWHM) of the RMSF, Faraday tomography may miss to reconstruct two FDF peaks for some specific intrinsic polarization angles. This phenomenon is known as RM ambiguities (\cite{far11,kum13}). Thus, Strategy B for unresolved sources is available if the gap caused by the IGMF is sufficiently larger than the FWHM of the RMSF. Such a limitation is practically relaxed if we include low frequency data providing a small FWHM of $O(0.1)~{\rm rad~m^{-2}}$. Thanks to progresses of low frequency radio observations such as LOFAR and Murchison Widefield Array (MWA), we would easily access low frequency data and could remove the ambiguity.

\subsection{FDF Model and Calculation}
\label{section2.6}

Instead of observational data, we use the data calculated from FDF models. We adopt simple FDF models, since the gap appears regardless of detailed profiles of FDFs for sources. The analyses below are rather independent on the shape of the FDF, because it is to be reconstructed by observations.

A FDF of a diffuse source is modeled with a function based on the hyperbolic tangent, since the function tends to better reproduce the resultant platykurtic profile expected from the Galaxy models (\cite{sun08,wae09,arkg13,ide14b}):
\begin{eqnarray}\label{eq:F_d}
F_{\rm d}(\phi)
&=& F_{\rm d0}
    \left[\frac{1}{4}
     \left\{\tanh{\left(\pi\frac{\phi - \phi_{\rm dw}}{\phi_{\rm dw}}\right)}
            + 1
     \right\}
    \right.
\nonumber \\
&\times&
\left.
 \left\{\tanh{\left(-\pi\frac{\phi - \phi_{\rm d} - \phi_{\rm dw}}
                             {\phi_{\rm dw}}
              \right)}
        + 1
 \right\}
\right],
\end{eqnarray}
where $\phi_{\rm dw}$ is the parameter which controls the width of the skirts, and $\phi_{\rm d}$ is the position of the back side edge of the FDF in $\phi$ space. We set $\phi_{\rm 0}=0$ for the position of the front side edge, where no generality is lost by this choice since the gap is measured from relative position of the sources in $\phi$ space. We choose $\phi_{\rm dw} = 2.0~{\rm rad~m^{-2}}$ and the thickness of the FDF $\phi_{\rm d} -\phi_{\rm 0} = 20~{\rm rad~m^{-2}}$ as representative values. We normalize FDF intensities of any sources by $F_{\rm d0}=1$, since relative amplitude of FDFs is only meaningful while we do not consider observational errors.

A FDF of a compact source such as a quasar and a radio galaxy has been approximated by a Gaussian (\cite{burn66,fssb11}):
\begin{equation}\label{eq:F_c}
F_{\rm c}(\phi)
= F_{\rm c0}
  \exp{\left\{-\frac{(\phi-\phi_{\rm c})^2}{2 \phi_{\rm cw}^2}\right\}},
\end{equation}
where $\phi_{\rm c}$ is the position of the FDF center in $\phi$ space, and $\phi_{\rm cw}$ determines the thickness of the FDF. We choose $\phi_{\rm cw}=0.2~{\rm rad~m^{-2}}$, and the thickness as $\pm 3\sigma$ region of the FDF is $2\times 3 \phi_{\rm cw} = 1.2~{\rm rad~m^{-2}}$ as a reachable value for selected sources (Sections 2.2 and 2.5). Such a value is actually expected for some sources toward high Galactic latitudes (e.g., \cite{sc86}). We investigate compact sources with $F_{\rm c0}=0.1-10000$.

A FDF of the cosmic web is modeled with
\begin{equation}\label{eq:F_IGM}
F_{\rm IGM}(\phi) = 0.
\end{equation}
This means that no emission is considered. As already mentioned, polarized intensities due likely to synchrotron radiations of cosmic-ray electrons in the IGM are generally very small except radio halos and relics in galaxy clusters. Faraday depth of the cosmic web is based on theoretical predictions (\cite{ar10,ar11}); the rms value of RM is $\sim 1~{\rm rad~m^{-2}}$ through a single filament at the local universe, and ${\rm several}-10~{\rm rad~m^{-2}}$ in integrating filaments up to $z=5$. \citet{ar11} also found that some of LOSs going through dense filaments and/or group of galaxies in filaments have RMs of a few tens ${\rm rad~m^{-2}}$. Therefore, we consider RM of the cosmic web, $1-30~{\rm rad~m^{-2}}$. Note that probing the RM of $O(1)~{\rm rad~m^{-2}}$ would be fundamentally difficult because of contaminations (sections \ref{section2.2} and \ref{section2.3}). But we also test the case to see the capability of Faraday tomography for such a small gap.

\begin{figure}[tp]
\begin{center}
\FigureFile(80mm,40mm){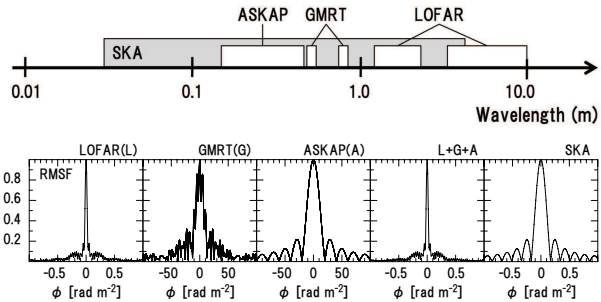}
\end{center}
\caption{Wavelength coverages and the RMSFs for LOFAR (L), GMRT (G), ASKAP (A) observations, combined observations of L+G+A, and the SKA observation.}
\label{f3}
\end{figure}

\begin{table}
\caption{Specifications of radio observatories.}\label{table1}
\begin{center}
\begin{tabular}{lccc}
\hline\hline
Observatory & Frequency & $\lambda^2$ & Channel \\
& (GHz) & (${\rm m^2}$) & \\
\hline
LOFAR LBA$^a$ & 0.030--0.080 & 14.00--99.00 & 62,000 \\
LOFAR HBA$^a$ & 0.120--0.240 & 1.600--6.200 & 156,000 \\
GMRT 327$^b$ & 0.305--0.345 & 0.760--0.970 & 256 \\
GMRT 610$^b$ & 0.580--0.640 & 0.220--0.270 & 256 \\
ASKAP$^c$ & 0.700--1.800 & 0.027--0.180 & 60,000\\
SKA$_2$ low$^d$ & 0.070--0.450 & 0.444--18.37 & 380,000 \\
SKA$_2$ mid$^d$ & 0.450--10.00 & 0.0009--0.444 & 67,000 \\
\hline
\multicolumn{4}{l}{$^a$LOFAR page; http://www.astron.nl/radio-observatory/} \\
\multicolumn{4}{l}{astronomers/lofar-astronomers} \\ 
\multicolumn{4}{l}{$^b$\citet{ana95}} \\
\multicolumn{4}{l}{$^c$ASKAP page; http://www.atnf.csiro.au/projects/askap/}\\
\multicolumn{4}{l}{$^d$SKA Phase 2, memo 130; http://www.skatelescope.org/} \\
\multicolumn{4}{l}{pages/page$\_$memos.htm}\\
\end{tabular}
\end{center}
\end{table}

The procedure of calculation is as follows. First, we construct the model FDF based on equations (\ref{eq:F_c}), (\ref{eq:F_d}) and (\ref{eq:F_IGM}). Next, we numerically carry out a Fourier transform of the FDF and synthesize the polarized intensity, $P(\lambda^2)$, according to Equation~(\ref{eq1}). After that, we derive the observable polarized intensity, $\tilde{P}(\lambda^2)$, by using the window function, $W(\lambda^2)$, in equation (\ref{eq:tildeP}), considering the observable bands of LOFAR, GMRT, ASKAP, and SKA listed in table \ref{table1} and shown in figure \ref{f3}. Finally, we numerically carry out a inverse Fourier transform of $\tilde{P}(\lambda^2)$ and obtain the reconstructed FDF, $\tilde{F}(\phi)$, according to equation (\ref{eq:tildeF}).

The model FDF is composed of $2.4\times 10^6$ data points ranging from $-1.5\times 10^4$ to $1.5\times 10^4~{\rm rad~m^{-2}}$ and dividing the $\phi$ space evenly. The same number of data points are adopted for the polarized intensity data dividing $\lambda^2$ space evenly. The data points provide wider $\lambda^2$ coverage and at least ten times higher $\lambda^2$ resolution than those in the observations, ensuring to minimize numerical errors in the calculation of the mock polarized intensity. We allow that each interferometer has different frequency resolution (number of channels). Here, the frequency resolution is worse than that of the generated data, so that the intensity in each channel is given by averaging the intensities of data points within each channel.

Hereafter, we do not take into account observational errors to demonstrate that the quality of reconstruction is mainly determined by frequency coverage. The cases with observational errors are discussed in Section \ref{section4}.

\section{Result}
\label{section3}

\subsection{Strategy A: Compact Source behind Diffuse Source}
\label{section3.1}

We first show the results for Strategy A. We define the RM of the gap caused by the cosmic web A as
\begin{equation}
{\rm RM_{IGMF}}=\phi_{f,B}-\phi_{d,A}.
\end{equation}

\begin{figure}[tp]
\begin{center}
\FigureFile(80mm,40mm){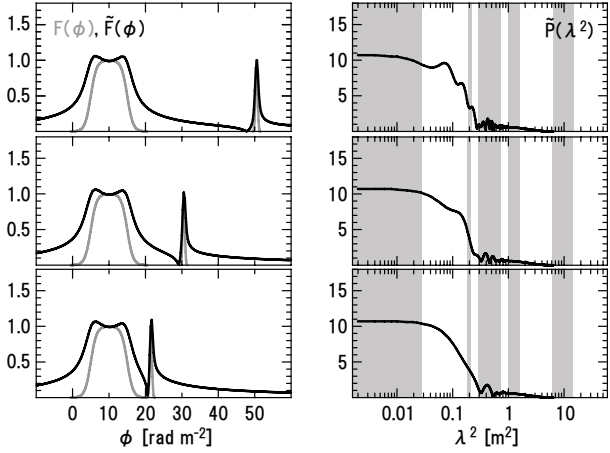}
\end{center}
\caption{Left panels show the model (gray) and reconstructed (black) Faraday dispersion functions. Right panels show synthesized polarized intensities, where unshadowed regions correspond to wavelength coverages with LOFAR, GMRT, and ASKAP. Panels from the top to bottom show the results for the models with $F_{\rm c0}/F_{\rm d0}=$ 1 and ${\rm RM_{IGMF}}=$ 30, 10, and 1 ${\rm rad~m^{-2}}$, respectively, for the SKA observation.}
\label{f4}
\end{figure}

Figure \ref{f4} shows model FDFs, synthesized polarized intensities, and reconstructed FDFs with the fixed intensity ratio, $F_{\rm c0}/F_{\rm d0}=1$, and ${\rm RM_{IGMF}}=$ 30, 10, and 1 ${\rm rad~m^{-2}}$, for SKA observations. We can see the gap due to the cosmic web, although its amplitude does not come to zero. 
This is because the reconstructed FDF of the diffuse source has skirts derived from the RMSF. A slope of the skirt changes at $\phi\sim 20$ ${\rm rad~m^{-2}}$ in the cases with ${\rm RM_{IGMF}}=$ 30 and 10 ${\rm rad~m^{-2}}$. If we regard this change as the edge of the diffuse source, we could estimate ${\rm RM_{IGMF}}$ from the gap within, at least, a factor of two. On the other hand, for ${\rm RM_{IGMF}}=$ 1 ${\rm rad~m^{-2}}$, it is not easy to recognize the change of the slope, and we need further analyses to remove skirts.

The polarized intensity decreases at the longer-wavelengths, $\lambda^2\gtrsim 0.1-1~{\rm m^2}$, at which Faraday depolarization (e.g., \cite{sok97}) happens because the Faraday rotation exceeds $\pi$ radian. In addition, the emission of the diffuse source interferes with that from the compact source, and results in further depolarization seen as wiggles in ASKAP and GMRT bands. Further general explanations for the behavior of polarized intensity can be obtained in BD05.

\begin{figure}[tp]
\begin{center}
\FigureFile(80mm,40mm){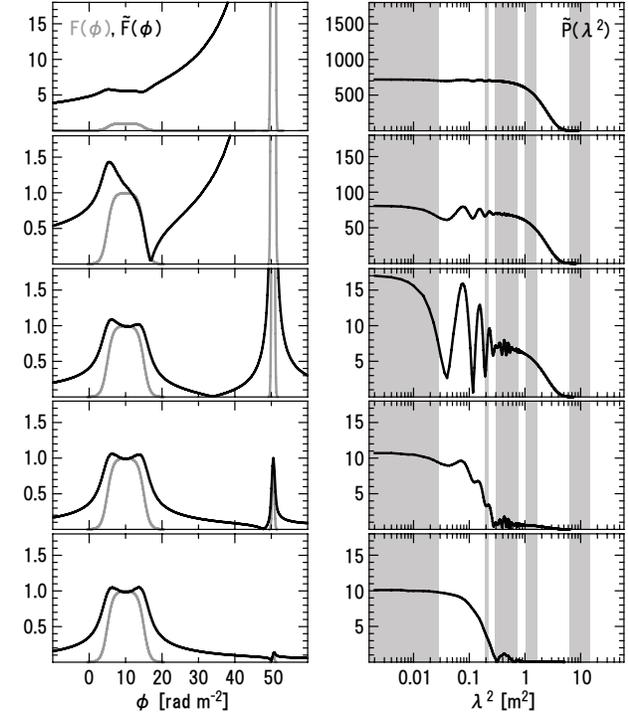}
\end{center}
\caption{Same as Figure \ref{f4} but for the models with ${\rm RM_{IGMF}}=$ 30 ${\rm rad~m^{-2}}$ and $F_{\rm c0}/F_{\rm d0}=$ 1000, 100, 10, 1, and 0.1, respectively, from the top to the bottom panels.}
\label{f5}
\end{figure}

We next change the intensity ratio, $F_{\rm c0}/F_{\rm d0}$. Figure \ref{f5} shows the results for ${\rm RM_{IGMF}}=$ 30 ${\rm rad~m^{-2}}$. Clearly, the promising range to find the gap is $F_{\rm c0}/F_{\rm d0}=1-10$. If the compact source is much brighter ($F_{\rm c0}/F_{\rm d0}\gtrsim 100$), the gap is substantially buried under the skirt extended from the compact source. On the other hand, if the compact source is much fainter ($F_{\rm c0}/F_{\rm d0}\lesssim 0.1$), it may become more difficult to identify the compact source from the skirt extended from the diffuse source.

The usefulness of the sources with $F_{\rm c0}/F_{\rm d0}=1-10$ can be also understood from behavior of the polarized intensity; we obtain strong wiggles in ASKAP and GMRT bands, which depend on ${\rm RM_{IGMF}}$ (Figure \ref{f4}). This means that observed polarized intensity has significant information about ${\rm RM_{IGMF}}$. Actually, if the compact source is too much brighter or too much fainter than the diffuse source, the polarized intensity has a simple monotonic form and the wiggles become faint.

It should be noticed that the reconstructed FDFs for $F_{\rm c0}/F_{\rm d0} \sim 100-1000$ show substructures around $\phi\sim 0-20$ ${\rm rad~m^{-2}}$, which cannot be produced only by a compact source. This indicates that the diffuse foreground emission is still significant, even if it is a few order fainter than the background emission. Actually, the wiggles which have information about ${\rm RM_{IGMF}}$ can be seen in the polarized intensity. Therefore, these cases would be also promising to find the gap, supported by further analyses to remove the skirt.

\begin{figure}[tp]
\begin{center}
\FigureFile(80mm,40mm){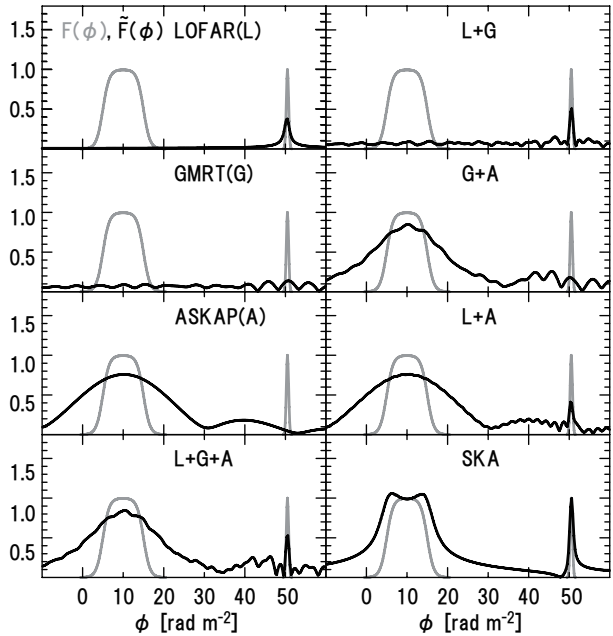}
\end{center}
\caption{Model (gray) and reconstructed (black) Faraday dispersion functions. Each panel shows the result for LOFAR (L), GMRT (G), ASKAP (A) observations, combined observations of L+G, G+A, A+L, L+G+A, and the SKA observation.}
\label{f6}
\end{figure}

Figure~\ref{f6} shows the reconstructed FDFs for the LOFAR, GMRT, ASKAP observations, their combinations, and the SKA observation, in the case with $F_{\rm c0}/F_{\rm d0}=1$ and ${\rm RM_{IGMF}}=30~{\rm rad~m^{-2}}$. The results clearly demonstrate that the quality of reconstruction mainly depends on frequency coverage of the data; fine RM structures with scales of $\lesssim {\rm a~few}~{\rm rad~m^{-2}}$ are mostly reconstructed with the low frequency data, while broad RM structures with scales of $\gtrsim {\rm several}~{\rm rad~m^{-2}}$ are mainly reconstructed with the mid-frequency data. Consequently, there exist skirts and side lobes derived from the RMSF, and we see a clear gap only for the SKA observation.

\subsection{Strategy B: Pair Compact Sources}
\label{section3.2}

\begin{figure}[tp]
\begin{center}
\FigureFile(80mm,40mm){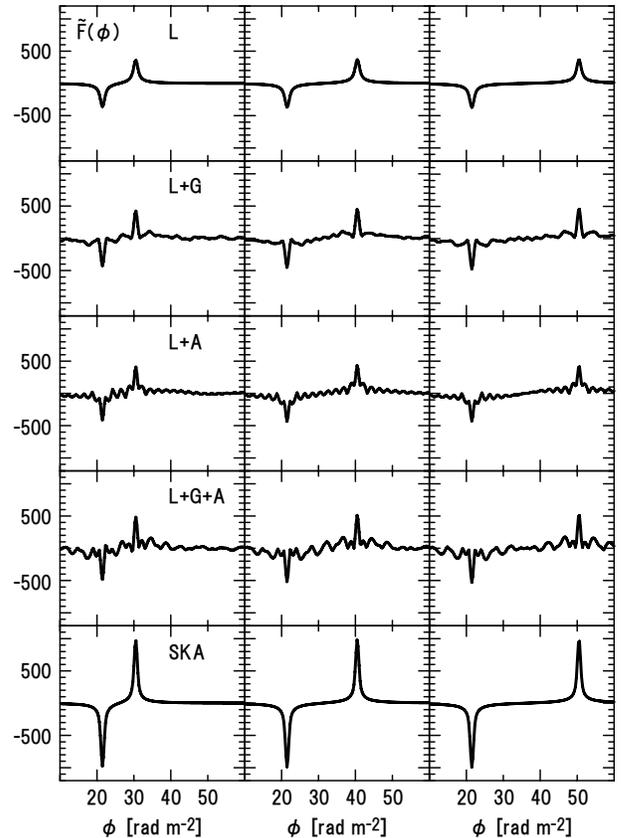}
\end{center}
\caption{Difference between two reconstructed Faraday dispersion functions for two LOSs. Panels from the top to bottom show the results for different observations, where the labels are the same as those in Figure~\ref{f6}. Panels from the left to right show the results with the RM of the cosmic web B, $\phi_{f,C}-\phi_{f,B}=$ 10, 20, and 30 ${\rm rad~m^{-2}}$ respectively, where $F_{\rm c0}/F_{\rm d0}=1000$.}
\label{f7}
\end{figure}

We next show the results for Strategy B. Figure~\ref{f7} shows the difference between two reconstructed FDFs for various combinations of telescopes and RMs of the cosmic web B, where we assumed a complete subtraction of the source A (the Galaxy). The two sharp peaks associated with the two sources can be reconstructed with the LOFAR observation, although the FDFs have skirts around the peaks and the peak intensities are underestimated by $\sim 60~\%$. The underestimation is slightly improved when we add GMRT and/or ASKAP data, but side lobes with an amplitude of at most $\sim 40~\%$ of the reconstructed peak arise. The side lobes could be ascribed to an absence of the data in $\lambda^2 \sim 0.3-0.7~{\rm m^{2}}$. The SKA observation nicely reproduces the FDF, though skirts extend to $4.5\sigma$ level with amplitudes of less than ${\rm several}~\%$ of the peak values. We find that the above features do not significantly change within $F_{\rm c0}/F_{\rm d0}=100-10000$. 

\begin{table}
\caption{Estimated ${\rm RM_{IGMF}}$ values.}\label{table2}
\begin{center}
\begin{tabular}{lccc}
\hline\hline
Actual value & 8.8 & 18.8 & 28.8 \\
& $({\rm rad~m^{-2}})$ & $({\rm rad~m^{-2}})$ & $({\rm rad~m^{-2}})$ \\
\hline
L$^a$ & 7.4 & 18.4 & 27.3 \\
LG$^a$ & 7.9 & 18.8 & 27.9 \\
LA$^a$ & 8.0 & 18.9 & 27.9 \\
LGA$^a$ & 7.9 & 18.0 & 27.9\\
SKA & 7.6 & 17.5 & 27.6 \\
\hline
\multicolumn{4}{l}{$^a$L: LOFAR, G: GMRT, A:ASKAP} \\
\end{tabular}
\end{center}
\end{table}

We define the RM of the gap caused by the cosmic web B as 
\begin{equation}
{\rm RM_{IGMF}}=\phi_{f,C}-\phi_{b,B},
\end{equation}
which give the lower-limit of the RM due to the cosmic web B, and the whole of the RM if $\phi_{\rm cw}\rightarrow 0$. In order to estimate ${\rm RM_{IGMF}}$, we refer to the interval of the two peaks seen in the FDF, since we confirmed that the relative Faraday depth between the two peaks are always precisely reconstructed. Here, the interval also includes skirts and side robes caused by the RMSF as well as ${\rm RM_{c,B}}/2+{\rm RM_{c,C}}/2$, where ${\rm RM_c}$ is the Faraday thickness of the compact source, i.e. $\phi_b-\phi_f$. Instead of estimating the Faraday thicknesses, we simply calculate a half width at half maximum of each peak, and subtract the half widths of the two peaks from the interval. The estimated ${\rm RM_{IGMF}}$ are listed in Table 2. We find that this method estimates ${\rm RM_{IGMF}}$ with errors less than $10-25~\%$ for all the cases shown in Figure~\ref{f7}. Note that the gap is substantially buried under the skirts in the case with ${\rm RM_{IGMF}}=1~{\rm rad~m^{-2}}\sim {\rm RM_c}$ (not shown), and ${\rm RM_{IGMF}}$ is not estimated correctly. Therefore, the above accuracy can be obtained, if ${\rm RM_{IGMF}}\gg{\rm RM_c}$.

\section{Discussion}
\label{section4}

\begin{figure}[tp]
\begin{center}
\FigureFile(80mm,40mm){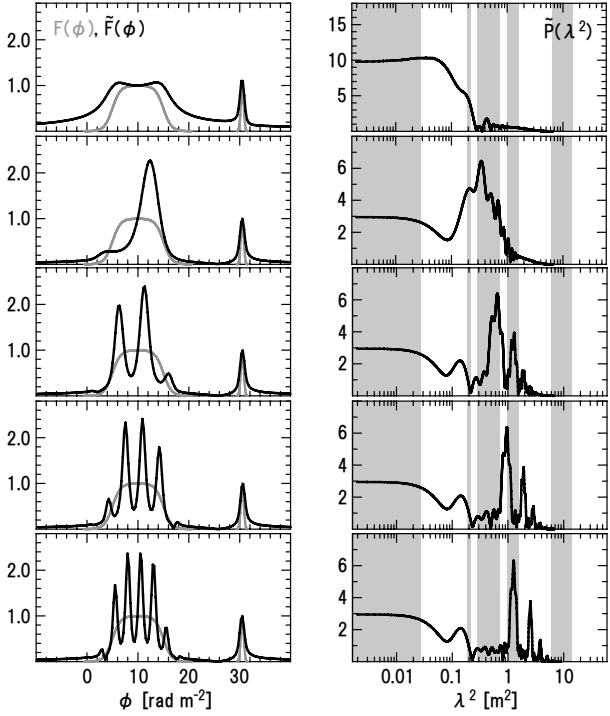}
\end{center}
\caption{Same as Figure \ref{f4} but for the models with $F_{\rm c0}/F_{\rm d0}=1$ and ${\rm RM_{IGMF}}=10~{\rm rad~m^{-2}}$ and $n=0,1,2,3,4$ (see text) for the SKA observation from the top to bottom panel, respectively.}
\label{f8}
\end{figure}

We have presented the cases for a pure real $F(\phi)$ obtained if the intrinsic polarization angle, $\chi_0$, does not depend on $\phi$. $\chi_0$ is, however, determined by structures of magnetic fields, and could be a function of $\phi$. In order to see effects of $\chi_0(\phi)$ on the reconstruction, we consider a variable $\chi_0(\phi)$ in our model. We multiply a phase factor $e^{2i \chi_0(\phi)}$ to a real function of $F(\phi)$ in Equation (\ref{eq1}), keeping the absolute value of the model FDF to be the same. For $\chi_0(\phi)$, although its general profile is not known, some characteristic behaviors could be understood by using a simple analytic function. We consider a periodic function $\chi_0(\phi) = \cos (2\pi \phi \times 0.1n)$ for $n=$ 1, 2, 3, and 4, since periodicity is expected from multiple reversals of turbulent magnetic fields. The results with $F_{\rm c0}/F_{\rm d0}=1$ and ${\rm RM_{IGMF}}=10$ ${\rm rad~m^{-2}}$ are shown in Figure \ref{f8}. We find that the profiles of the reconstructed FDFs highly depend on $n$. Nevertheless, the edges of the sources are rather sharp compared with the fiducial model ($n=0$). This may be ascribed to the cancellation of polarized emissions at the tails due to a rotation of the intrinsic polarization angle. Therefore, our main simulations could be regarded as conservative cases with largest extension of the skirts, which would be somewhat reduced in realistic situations. Eventually, ${\rm RM_{IGMF}}$ could be better estimated from the gap between the two sources.

A real FDF of the Galaxy would be much more complex. It would have $n\gg 4$ based on the coherence length of magnetic fields of several tens of pc (see \cite{arkg13}, references therein). Even in such a case, our observational strategies would be still available, since the intrinsic polarization angle does not alter the key feature: two sources and the gap between them (Figure \ref{f8}). Multiple peaks may become an ambiguity for identifying which peak is an extragalactic origin and which gap is caused by the IGMF. But we could solve the ambiguity, if we carry out an ``off-source" observation and gain the FDF of the Galaxy only.

We notice that a real FDF of the Galaxy should depend on Galactic longitude and latitude as well as properties of turbulent magnetic fields such as the driving scale, the Mach number, the plasma $\beta$, and so on (\cite{arkg13}). Although considerations of them are beyond the scope of this paper, developing realistic FDFs of the Galaxy must be an important subject to make the detection of the gap more reliable. Realistic FDFs of the Galaxy based on Akahori et al. (2013) will be presented in a separate paper (\cite{ide14b}).

\begin{figure}[tp]
\begin{center}
\FigureFile(80mm,40mm){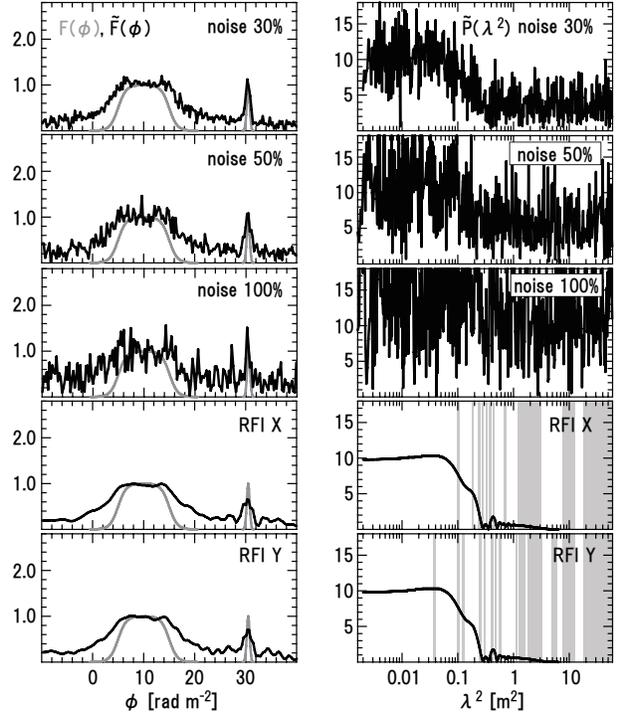}
\end{center}
\caption{Same as Figure \ref{f4} but for the models with $F_{\rm c0}/F_{\rm d0}=1$ and ${\rm RM_{IGMF}}=10~{\rm rad~m^{-2}}$ and with observational effects: top three panels are for the cases with noise amplitudes of 30, 50, and 100 \% of the polarized intensities, and the bottom two panels are for the cases with RFIs on sites X and Y, where unshadowed regions are wavelength coverages for SKA without strong RFIs.}
\label{f9}
\end{figure}

Another simplification in this paper was to neglect observational effects. Particularly, it is true that there are significant noise on polarized intensities and some frequencies are probably missing due to radio frequency interferences (RFIs). We demonstrate these effects as follows.

\begin{figure}[tp]
\begin{center}
\FigureFile(80mm,40mm){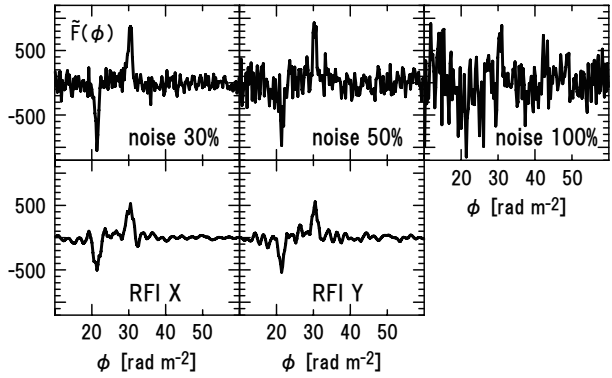}
\end{center}
\caption{Same as Figure \ref{f7} but for the models with the RM of the cosmic web B, $\phi_{f,C}-\phi_{f,B}=$ 10 ${\rm rad~m^{-2}}$ and with observational effects: top three panels are for the cases with noise amplitudes of 30, 50, and 100 \% of the polarized intensities, and the bottom two panels are for the cases with RFIs on sites X and Y.}
\label{f10}
\end{figure}

For effects of observational noise, we include the noise into the observable polarized intensity, $\tilde{P}(\lambda^2)$, and get the reconstructed FDF. We consider a random Gaussian noise in each $\lambda^2$ domain. The results with noise amplitudes of 30, 50, and 100 \% of the polarized intensities for representative cases are shown in top panels of Figures \ref{f9} and \ref{f10}. We see that the noise amplitude of 30 \% does not dramatically alter the overall profile of the FDF, and the reconstructed FDF would be useful up to the noise amplitude of $\sim 50~\%$. Such a requirement of the noise would limit the sample of radio sources that could be considered.

As for effects of RFIs, we discard the data in frequencies where significant RFIs exist, and get the reconstructed FDF. We refer to the recent assessment report\footnote{http://www.skatelescope.org/wp-content/uploads/2012/06/78g\_SKAmon-Max.Hold\_.Mode\_.Report.pdf} of RFIs for the SKA candidate sites, X and Y. The results for representative cases are shown in bottom panels of Figures \ref{f9} and \ref{f10}. We can see less-peaked profiles for compact sources due to lack of data in low frequencies around $\sim 86-108$ MHz and $\sim 170-270$ MHz. Such a broadening of the FDF for the compact source would produce uncertainties of a few ${\rm rad~m^{-2}}$ on the estimation of the gap.

Reconstructed FDFs have significant skirts and side lobes originating from the RMSF. Such skirts and side lobes are a major ambiguity to probe the IGMF. The issues related to the RMSF could be, however, improved by using decomposition techniques such as RMCLEAN \citep{hea09} and calibrations of the RMSF by phase correction (BD05) and symmetry assumption \citep{fssb11}. Also, wavelet-based tomography \citep{fssb11,bfss12} would allow better representation of localized structures in the data unlike with decompositions with harmonic functions in the Fourier transform. QU-fitting (\cite{osu12,ide14a}) and compressive sampling/sensing \citep{don06,ct06,lbcd11,ast11} would be also promising to probe the gap caused by the IGMF.

Another important improvement to get better reconstructions of the FDF is even sampling in $\lambda^2$ space. Although we have assumed even sampling in the simulations, observations sample the data evenly in $\lambda$ space so far. Such data produce unevenly-sampled data in $\lambda^2$ space, and cause large numerical artifacts in Fourier transform. In order to minimize numerical errors on the Fourier transform, development of flexible receiver systems which allow us to sample the data evenly in $\lambda^2$ space would be a key engineering task for future radio astronomy (e.g., CASPER/ROACH\footnote{https://casper.berkeley.edu/wiki/ROACH}).

\section{Summary}
\label{section5}

In order to probe Faraday rotation measure (RM) of the intergalactic magnetic field (IGMF) in the cosmic web from observations of RMs for extragalactic radio sources, we need to separate contributions of other origins of RMs. In this paper, we discussed possible observational strategies to estimate the RM due to the IGMF by means of Faraday tomography (Faraday RM Synthesis). Our quantitative discussion indicated that there are two possible strategies - an observation of a compact source behind a diffuse source (Strategy A) and a comparison between the observation with another observation for a nearby compact source (Strategy B).

For the two strategies, we investigated the capability of Faraday tomography in present and future wide-band radio polarimetry. For Strategy A, we confirmed that the RM due to the IGMF can be seen as the relative Faraday depth between the two sources. A promising polarized intensity of the compact source relative to the diffuse source is $\sim 1-10$. As for Strategy B, we found that the relative Faraday depth of the compact sources gives a reasonable estimate of the RM with errors less than a few tens percents for LOFAR or SKA observations, if the RM is larger than $10~{\rm rad~{m^{-2}}}$. Such an accuracy is expected while the compact sources are $\sim 100-10000$ times brighter than the diffuse source. Strategy B provides better estimations of the RM, but Strategy A is also important to increase the chance of the estimation.

Since we have adopted simple models of radio sources, we discussed more realistic cases for specific situations. We demonstrated that the multiple changes of intrinsic polarization angle within a diffuse source would not make the estimation of the relative Faraday depth between the two sources worse. Multiple peaks may become an ambiguity for identifying the gap caused by the IGMF, but the ambiguity could be solved with an off-source observation. We also considered effects of observational noise and radio frequency interferences, and found that these effects can be practically insignificant for the estimation, at least for the moderate levels of noise and RFI considered here.

Our simulations and discussions indicate that it is still not easy to explore RM of $O(1)~{\rm rad~m^{-2}}$ for the IGMF, because of incompleteness of the reconstruction as well as ambiguities due to RMs associated with the source and intervening clouds. These issues would be, however, improved by decomposition techniques as well as theoretical and observational studies of Faraday depolarization. With the improvements, we would finally reach observations of the IGMF in filaments of galaxies.

\vskip 11pt

The authors would like to thank the referee(s) for constructive comments and discussions, B. M. Gaensler, S. P. O'Sullivan, and X. H. Sun for useful comments. T.A. and K.K. acknowledge the supports of the Japan Society for the Promotion of Science (JSPS). T.A. and D.R. were supported by National Research Foundation of Korea through grant 2007-0093860. K.T. was supported by Grants-in-Aid from MEXT of Japan, No.~23740179, No.~24111710 and No.~24340048.

\end{document}